\documentclass[10pt,a4paper]{article}
\usepackage{multicol}
\usepackage{graphicx}
\usepackage{authblk}
\usepackage{float}
\usepackage{amsmath,amssymb,amsfonts}
\usepackage[width=18cm,height=24cm]{geometry}
\usepackage{hyperref}

\newcommand \med[1] {\langle{#1}\rangle}
\newcommand {\vk} {\bf k}
\newcommand \vri {{{\bf R}}_{{\textnormal i}}}
\newcommand \ves {{\bf S}}
\newcommand \sa {\sigma}
\newcommand \so {\bar{\sigma}}
\newcommand \up {\uparrow}
\newcommand \pu {\downarrow}
\newcommand \al {\alpha}

\newcommand \be {\beta}
\newcommand \lam[3] {\lambda_{{#1}{#2}{#3}}}
\newcommand \lan[3] {\lambda_{{#1}{#2}{#3}}^{\dag}}
\newcommand \lws[1] {\med{\lambda_{#1}}}
\newcommand \epf {\epsilon_f}
\newcommand \epk[1] {\epsilon_{{\bf #1}}}
\newcommand \ckm[2]  {c_{{\bf #1}{#2}}^{\dag}}
\newcommand \ck[2]     {c_{{\bf #1}{#2}}}

\newcommand \fim[3] {f_{{#1}{#2}{#3}}^{\dag}}
\newcommand \fis[3] {f_{{#1}{#2}{#3}}}
\newcommand \zer   {|0\rangle}

\newcommand \ze[1] {|{#1}\rangle}
\newcommand \ez[1] {\langle{#1}|}
\newcommand \mzer {\big(f_{i1\up}^{\dag}f_{i2\pu}^{\dag}+
f_{i1\pu}^{\dag}f_{i2\up}^{\dag}\big)\zer}

\newcommand \vko[2] {V_{{\bf #1}\,{#2}}}
\newcommand \vka[2] {V_{{\bf #1}\,{#2}}^*}

\hypersetup{
pdfauthor = {Christopher Thomas},
pdftitle = {Draft anisotropy effect in the ukl model},
}

\begin{document}

\title{Effect of anisotropy  in the $S=1$ underscreened Kondo lattice}

\author[1,2]{Christopher Thomas\thanks{chris@if.ufrgs.br}\thanks{Present address: Instituto de F\'isica, UFRGS, 91501-970 Porto Alegre-RS, Brazil}}
\affil[1]{Institut de Physique Theorique, CEA-Saclay, 91191 Gif-sur-Yvette, France}
\affil[2]{International Institute of Physics, UFRN, 59078-400 Natal-RN, Brazil}
\author[3]{Acirete S. da Rosa Sim\~oes}
\affil[3]{Instituto de F\'isica, UFRGS, 91501-970 Porto Alegre-RS, Brazil}
\author[4]{Claudine Lacroix}
\affil[4]{Institut N\'eel, Universit\'e Grenoble-Alpes, F-38042 Grenoble, France and Institut N\'eel, CNRS, F-38042 Grenoble, France}
\author[3]{Jos\'e Roberto Iglesias}
\author[5]{Bernard Coqblin\thanks{Our colleague and friend Bernard Coqblin passed away on May 29, 2012 during the final stages of this work.}}
\affil[5]{Laboratoire de Physique des Solides, CNRS - Universit\'e Paris-Sud, F-91405 Orsay, France}

\maketitle

\begin{abstract}
We study the effect of crystal field anisotropy in the underscreened  $S=1$ Kondo lattice model. Starting from the two orbital Anderson lattice model and including a local  anisotropy term, we show, through Schrieffer-Wolff transformation, that local anisotropy is equivalent to an anisotropic Kondo interaction ($J_{\parallel} \neq{J_{\perp}}$). The competition and coexistence between ferromagnetism and Kondo effect in this effective model is studied within a generalized mean-field approximation. Several regimes are obtained, depending on the parameters, exhibiting or not coexistence of magnetic order and Kondo effect. Particularly, we show that  a re-entrant Kondo phase at low temperature can be obtained. We are also able to describe phases where the Kondo temperature is smaller than the Curie temperature ($T_K<T_C$). We propose that some aspects of uranium and neptunium compounds that present coexistence of Kondo effect and ferromagnetism, can be understood within this model.
\end{abstract}

\begin{multicols}{2}

\section{Introduction}

The properties of many cerium or ytterbium compounds are well accounted for by the $S=1/2$ Kondo-lattice model, where a strong competition exists between the Kondo effect and magnetic ordering arising from the RKKY (Ruderman-Kittel-Kasuya-Yosida) interaction between rare-earth atoms at different lattice sites. This situation is well described by the Doniach diagram,\cite{Doniach1977,Iglesias1997} which gives the variation of the N\'eel temperature and of the Kondo temperature with increasing antiferromagnetic intrasite exchange interaction $J_K$ between localized spins and conduction-electron spins. However, uranium and neptunium compounds exhibit a different behavior. The Kondo behavior of rare earth and actinide compounds depends on the number of $f$ electrons. For example, in the case of cerium and ytterbium rare earth systems, the localized $S=1/2$ spins of the $4f$ electrons interact with the spin of the conduction electrons via the $s-f$ exchange, leading to  Kondo and magnetic interactions that can be  described by the usual $S=1/2$  Kondo lattice model. The situation is different for the uranium and neptunium compounds. In this case,  the total spin of the $5f$ electrons is $S>1/2$ and screening may be only partial  when the number of conduction electrons channels $n$ is smaller than  $2S$~\cite{Nozieres1980}.  Many actinide compounds, like UTe~\cite{Schoenes1996}, UCu$_{0.9}$Sb$_2$~\cite{Bukowski2005}, NpNiSi$_2$~\cite{Colineau2008} and Np$_2$PdGa$_3$~\cite{Tran2010}, have been reported to exhibit an underscreened Kondo effect; all of them also exhibit ferromagnetic order, with  a  relatively large Curie temperature of the order of 50-100 K. Very recently it was found another neptunium compound, Np$_2$PtGa$_3$~\cite{Tran2014}, presenting a similar effect, i.e. coexistence of Kondo effect and ferromagnetism, but with a smaller Curie temperature $\sim26$ K.

Another important point is that the $5f$ electrons are less localized than the $4f$ electrons~\cite{Zwicknagl2002,Zwicknagl2003,Hoshino2013} and their localized character can be easily changed under pressure. For example, when applying pressure in UTe samples, the Curie temperature, $T_C$ initially increases, then reaches a maximum and decreases~\cite{Schoenes1984}. This can be understood as a decrease of the degree of localization of the $5f$ electrons under pressure~\cite{Sheng1996,Thomas2011}.

In previous works~\cite{Thomas2011,Perkins2007,Thomas2012}, the possible coexistence of ferromagnetic order and Kondo effect was discussed in detail. Particularly, the Schrieffer-Wolf transformation was applied to a degenerate periodic Anderson Hamiltonian with two  $f-$electrons on a degenerate $f-$level, resulting in a $S=1$ spin in the ground state, and a Kondo interaction between the $S=1$ spins and the conduction electrons was obtained, together with an effective $5f$ bandwidth that corresponds to a delocalization of the $f-$electrons~\cite{Thomas2011}. In this way it was possible to obtain a qualitative phase diagram that agrees with the experimental results for some uranium and neptunium compounds which show coexistence of Kondo effect and ferromagnetism.

Here we are interested in the effect of crystal field anisotropy on the $S=1$ Kondo lattice model. The effect of anisotropy has been well studied for the case of $S=1/2$ Kondo compounds, particularly cerium compounds~\cite{Cornut1972,Coqblin2006} where the specific heat and transport properties have been well understood by considering anisotropy. In the case of actinide compounds with larger values of the spin, and stronger delocalization  of the $f-$electrons,  we also expect that the behavior of the Kondo and Curie temperatures will change when anisotropy is considered. So, in the following we analyze  the anisotropic underscreened Kondo lattice (UKL) model, focusing on two different regimes which are peculiar to the $S=1$ UKL model : the re-entrant Kondo regime and the ferromagnetic phase with $T_K<T_C$.

\section{The UKL Model with anisotropy}

Following ref.~\cite{Thomas2011} we first describe the $5f$ electrons system using the periodic Anderson lattice model with two localized orbitals. This allow us to describe local $S=1$ magnetic sites. The Hamiltonian is written as follows:
\begin{align}\label{eq:UAL}
H=H_0+H_{hyb}\,,\quad H_0=H_s+H_f+H_{ani}\,,
\end{align}
where
\begin{align}
&H_{s}= \sum_{\vk\sa}\epk{k}\ckm{k}{\sa}\ck{k}{\sa}~,\\
&H_{f}=\sum_{i\al\sa}\epf n^{f}_{i\al\sa}+\sum_i\Big[U(n_{i1\up}^f n_{i1\pu}^f + n_{i2\up}^f n_{i2\pu}^f)\notag \\ & \hspace{-0.1cm}+U'(n_{i1\up}^fn_{i2\pu}^f+n_{i1\pu}^fn_{i2\up}^f)+(U'-J)(n_{i1\up}^fn_{i2\up}^f+n_{i1\pu}^fn_{i2\pu}^f)\nonumber\\
&-J(\fim{i}{1}{\up}\fis{i}{1}{\pu}\fim{i}{2}{\pu}\fis{i}{2}{\up}
+ h. c.)\Big]~, \\
&H_{hyb}=\sum_{i{\vk}\al\sa} \big(V_{\vk\al}e^{i{\vk}\cdot\vri}\ckm{k}{\sa}\fis{i}{\al}{\sa}+V^{*}_{\vk\al}e^{-i{\vk}\cdot\vri}\fim{i}{\al}{\sa}\ck{k}{\sa}\big),\\
&H_{ani}=-D\sum_{i}(S^f_z)^2_i\,.\label{eq:ani_d}
\end{align}
 $H_s$ represents the kinetic energy for the conduction electrons, being $\epk{k}$ the dispersion relation and $\ckm{k}{\sa}$ ($\ck{k}{\sa}$) the creation (annihilation) operator for conduction electrons. As usual, no degeneracy of the conduction band is considered. The local energy $\epf$, the Coulomb interaction of two $f$ electrons in the same orbital, $U$, in different orbitals, $U'$  and the Hund coupling, $J$, are included into $H_f$, with $\fim{i}{\al}{\sa}$ ($\fis{i}{\al}{\sa}$) being the creation (annihilation) operator in the site $i$, orbital $\al$ and spin $\sa$. Finally, the term $H_{hyb}$ represents the hybridization between the conduction electrons and the $f$ electrons and $H_{ani}$ is a local magnetic anisotropy, leading to uniaxial (for $D>0$) or planar anisotropy (for $D<0$).

This model has been partially studied in ref.~\cite{Thomas2011} but without the last term. Here we study the effect of  anisotropy. For this purpose we generalize the Schrieffer-Wolff (SW) transformation~\cite{Schrieffer1966} for the  Hamiltonian described in equation \eqref{eq:UAL} (details of the SW transformation are given in Appendix \ref{sec:ap_sc}). The anisotropy term in equation \eqref{eq:ani_d} removes the degeneracy between the $f$ states $S_z=|\pm1\rangle$ and $S_z=|0\rangle$. For positive values of the parameter $D$, the states with $S^f_z=\pm1$ have smaller energy than the one with $S_z=0$.

From the SW transformation of the Hamiltonian \eqref{eq:UAL}, the Kondo interaction can be written as:
\begin{align}
H_K=\frac{1}{2}\sum_{i\vk}&
\left[J_{\perp}(\ckm{k'}{\up}\ck{k}{\pu}S_i^{f-}+\ckm{k'}{\pu}\ck{k}{\up}S_i^{f+})\right. \notag \\
&\left.+J_{\parallel}(\ckm{k'}{\up}\ck{k}{\up}-\ckm{k'}{\pu}\ck{k}{\pu})S_{zi}^{f}\right]\,,
\end{align}
where the parallel (diagonal) and the perpendicular (non-diagonal) Kondo interactions are
\begin{align}\label{eq:param_ani_ja}
J_{\parallel}&=-2\frac{|V_{k_F}|^2}{U'-J+\epf-\mu-3D/4}, \notag\\
J_{\perp}&=-|V_{k_F}|^2\Big(\frac{1}{U'-J+\epf-\mu-3D/4}\notag \\&
+\frac{1}{U'-J+\epf-\mu+D/4}\Big)\,,
\end{align}
To stabilize ferromagnetic order, we include in the UKL model a ferromagnetic exchange interaction between neighboring spins $H_H=\frac{1}{2}J_{H}\sum_{\med{ij}} \ves^f_i\cdot\ves^f_j$, with $J_H<0$.

The presence of the anisotropy is explicitly visible in the expressions of these interaction parameters $J_{\parallel}$ and $J_{\perp}$, given by equation \eqref{eq:param_ani_ja}. Thus, the magnetic anisotropy induces an anisotropic Kondo interaction. It can be checked easily from equation \eqref{eq:param_ani_ja} that,  $D>0$ implies that $J_{\parallel}<J_{\perp}$. We define new parameters $\al$ and $J_K$ as follows:
\begin{align}
\alpha=\frac{J_{\parallel}}{J_{\perp}}\,\,,\text{and}\,\,\, J_K=\frac{J_{\parallel}+J_{\perp}}{2}\,.
\end{align}
 $D>0$ is equivalent to $\al<1$. 
The anisotropic $S=1/2$  Kondo lattice model has been studied by several authors  ~\cite{Zhang2000} without any microscopic justification of the anisotropy of the interaction;  here we show that for $S=1$, the anisotropic Kondo interaction results from the local anisotropy.

\section{Influence of anisotropy on the phase diagram}

In the following, we use the same generalized mean field decoupling scheme already utilized in refs.~\cite{Thomas2011,Perkins2007} and calculate self-consistently  the order parameters: $\med{M^f_i}=\med{S_i^z}\equiv\med{S_{i1}^z+S_{i2}^z}$ for the magnetization of the $f$ electrons and $\med{\lambda_{i\sa}}=\sum_{\al}\med{\fim{i}{\al}{\sa}c_{i\sa}}$ for the Kondo hybridization as a function of $J_K$, temperature and $\al$. From the variation of the critical temperature for ferromagnetic order, $T_C$, and  Kondo temperature, $T_K$, we can build the phase diagram. As it will be shown in the next paragraph, it is also necessary to introduce two characteristic temperatures $T_1$ and $T_1'$, associated with  re-entrant phases that appear when the order parameters, $\med{\lambda_{\sa}}$ and $\med{M^f}$, are zero at $T=0$ but different from zero above $T_1$ and $T_1'$.
In all the following calculations, the half bandwidth $W_D$ is taken as the energy unit, and we fix the intersite exchange $J_H =-0.01$, the conduction band filling to $\med{n^c}=0.8$, and the total number of $f$ electrons per site and per orbital to $\med{n^f}=1$.

In the UKL model, the $f$ electrons are not purely localized and a small effective bandwidth was obtained from the SW transformation~\cite{Thomas2011}. This  effective $f$ bandwidth is defined as
\begin{align}
W^f=2A_{\sa}\epk{k}\,,
\end{align}
where $A_{\sa}$ is proportional to the magnetization of the $f$ electrons $\med{M^f}$, the Kondo exchange interactions $J$'s and a numerical parameter $P$ (see equation \eqref{eq:aks} in Appendix \ref{sec:meanfield} and ref.\cite{Thomas2011}) that measures the width of the $f-$band compared to the width of the conduction band. Here, we consider $P$ as a free parameter, and we describe the results for two values of $P$: $P=0.20$ and for $P=0.30$.

\begin{figure}[H]
\centering
\includegraphics[width=1\columnwidth]{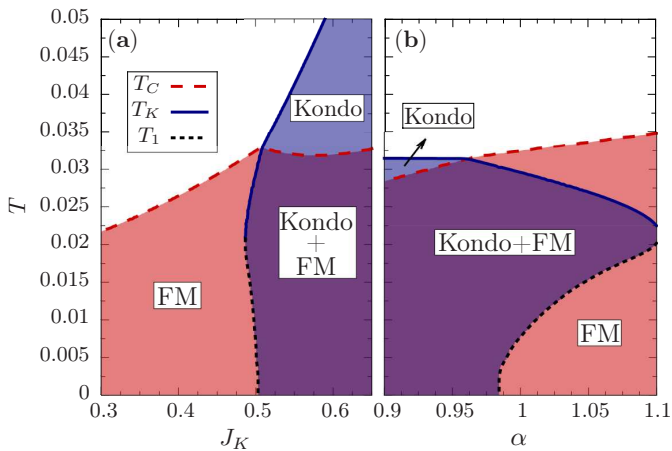}
\caption{\label{fig:ph_p020}({\it color online}) Phase diagrams for $P=0.20$. ({\bf a}) $T$ {\it versus} $J_K$ for $\alpha=1$ and ({\bf b}) $T$ {\it versus} $\alpha$ for $J_K=0.50$. The red color region represents the ferromagnetic order and the blue color region represents the Kondo phase. A region of coexistence can be observed in both phase diagrams.}
\end{figure}

\begin{figure}[H]
\centering
\includegraphics[width=1\columnwidth]{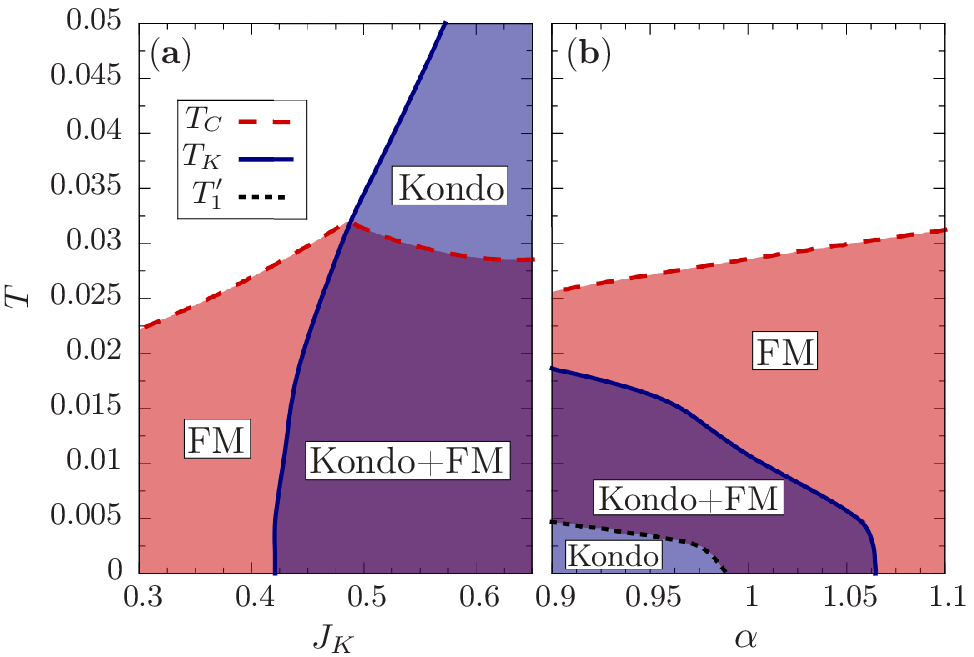}
\caption{\label{fig:ph_p030} ({\it color online}) Phase diagrams for $P=0.30$. ({\bf a}) $T$ {\it versus} $J_K$ for $\alpha=1$ and ({\bf b}) $T$ {\it versus} $\alpha$ for $J_K=0.43$. The red color region represents the ferromagnetic order and the blue color region represents the Kondo phase. A region of coexistence can be observed in both phase diagrams.}
\end{figure}

\begin{figure*}{hb}
\centering
\includegraphics[width=0.8\textwidth]{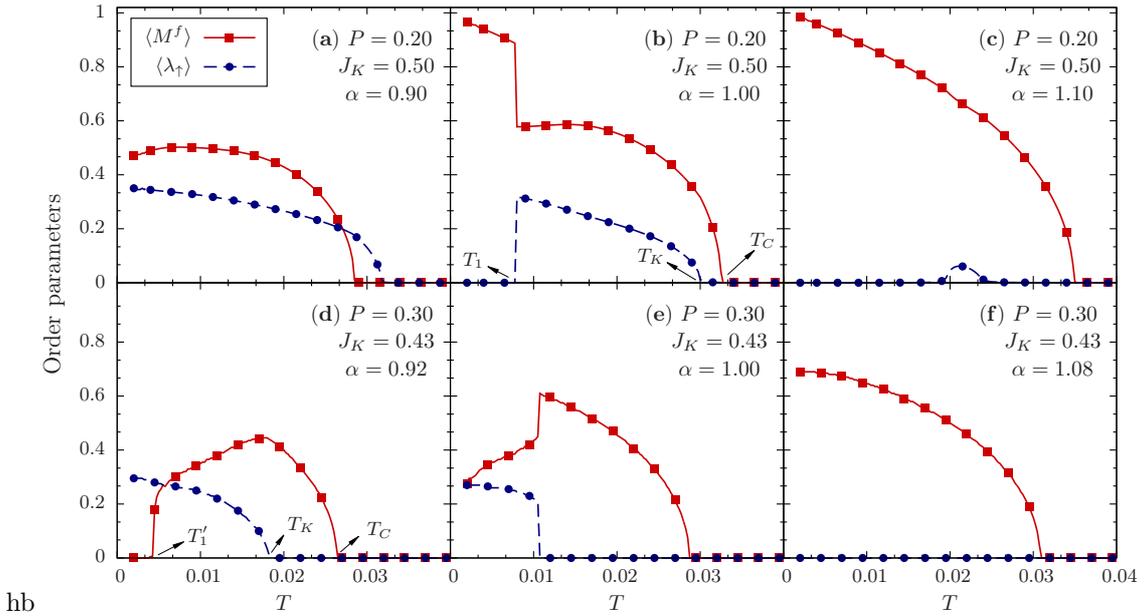}
\caption{\label{fig:pa_p020_p030} Variation of $\med{M^f}$ and $\med{\lambda_{\up}}$ as a function of temperature for different values of $\alpha$ in the two cases studied here, $P=0.20$ and $P=0.30$.}
\end{figure*}

The phase diagram as a function of $J_K$ is shown on figure \ref{fig:ph_p020}{(\bf a}) for $P=0.20$, and in the absence of anisotropy ($\alpha=1$). When the Kondo coupling $J_K$ is not strong enough ($J_K\lesssim0.51$),  Kondo effect is not stable for $T\rightarrow0$, but it is stable in a narrow temperature region for $J_K$ very close to 0.50. In this region, the Kondo effect can occur between two characteristic temperatures: $T_1$ and $T_K$. Figure \ref{fig:pa_p020_p030} shows the behavior of the mean field parameters as a function of temperature. Such re-entrant behavior corresponds to the case presented in figure \ref{fig:pa_p020_p030}({\bf b}) for $J_K=0.50$. Also, figure \ref{fig:ph_p020}{(\bf b)} shows the phase diagram for a value of $J_K=0.50$, and as a function of $\alpha$. The re-entrance is found in a large range of $\alpha$. The width of the re-entrance region is maximum for $\alpha \sim 1$ and decreases when $\alpha$ increases. On the other hand, the Kondo effect is enhanced  when $\alpha$ decreases to values smaller than $1$. Thus, it seems clear that the anisotropy plays against the Kondo effect while the Curie temperature is almost insensitive to changes in the anisotropy.

Figure \ref{fig:ph_p030}({\bf a}) shows the phase diagram for a larger  $f-$bandwidth, ($P=0.30$) and no anisotropy ($\alpha=1$). In this case, there is a region where $T_K<T_C$, i.e. Kondo effect start to occur within the ferromagnetic phase. Such a behavior corresponds to some experiments in actinide compounds~\cite{Colineau2008,Tran2010}. It can be seen on figure \ref{fig:pa_p020_p030}({\bf e}), close to the critical $J_K$ value below which Kondo effect disappears, that the transition is abrupt between a ferromagnetic state and mixed phase Kondo+ferromagnetic with a discontinuous jump of the magnetization. The variation of the critical temperatures for fixed $J_K=0.43$, as a function of anisotropy, is presented in figure \ref{fig:ph_p030}({\bf b}). In this case, there is also a re-entrant phase, but for the ferromagnetic phase: the ferromagnetic state is stable only between two temperatures $T_1'$ and $T_C$, but below $T_1'$, Kondo effect dominates and ferromagnetism disappears. 
This behavior can be clearly seen on figure \ref{fig:pa_p020_p030}({\bf d}), where it is evident that ferromagnetism disappears at low temperature.

Summarizing, for both values of $P$, $T_C$ has a similar behavior. It increases with increasing $J_K$ and has a maximum when $T_C$ is of the same order than $T_K$. This variation of $T_C$  corresponds to what was observed in experiments ~\cite{Schoenes1996,Link1992,Cornelius1996} and  it was already discussed from a theoretical point of view in~\cite{Cooper1998}, where it was proposed that this maximum is due to delocalization of the $5f$ electrons. Here, as in  our previous papers, ~\cite{Perkins2007,Thomas2011}, we propose that this maximum is related to the competition with Kondo effect.

Figure \ref{fig:pa_p020_p030} presents the variation of the order parameters $\med{M^f}$ and $\med{\lambda_{\up}}$ as a function of  temperature, for different values of $\al$ for  $P=0.20$, (figures \ref{fig:pa_p020_p030}({\bf a}-{\bf c})), and $J_K=0.50$, and for $P=0.30$ and $J_K=0.43$, (figures \ref{fig:pa_p020_p030}({\bf d}-{\bf f})). The values of $\alpha$ in figure \ref{fig:pa_p020_p030} were chosen to emphasize the difference between the possible scenarios present for this model.

For $P=0.20$, the coexistence of FM and Kondo effect is clearly observed for $\al=0.90$, where $T_K>T_C$ (figure \ref{fig:pa_p020_p030}({\bf a})). Close to the case without anisotropy $(\al\approx1)$, the behavior is very interesting, as already pointed out: at low $T$, the system presents a purely FM phase, while a coexistence FM + Kondo is observed above a temperature $T_1$ (figure \ref{fig:pa_p020_p030}({\bf b})). Increasing $\al$, ferromagnetism dominates, and  Kondo effect is observed only in a small temperature-region (figure \ref{fig:pa_p020_p030}({\bf c})). On the other hand, the results for $P=0.30$ are very different. For $\al\lesssim0.96$,  Kondo effect is strong enough to destroy completely the FM state at low temperature (figure \ref{fig:pa_p020_p030}({\bf d})). A coexistence of FM + Kondo is observed at low $T$ when $\al\approx1$ (figure \ref{fig:pa_p020_p030}({\bf e})). With decreasing temperature, there is first a ferromagnetic state between $T_C$ and $T_K $ and then a mixed Kondo-ferromagnetic state below $T_K$ with a clear decrease of the ferromagnetic magnetization. In the Kondo-ferromagnetic phase, the $f$-moments are partially screened, and the $f$-moments are reduced to a value close to $1/2$. This is in agreement with the description given by Nozi\`eres and Blandin \cite{Nozieres1980} of the underscreened Kondo effect: the conduction electrons cannot completely screen a $S=1$ localized spin, and the system is described below $T_{K}$ by an effective $S=1/2$ spin, which can order due to the intersite exchange. For bigger values of $\al$ the FM is predominant and no Kondo effect is observed (figure \ref{fig:pa_p020_p030}({\bf f})).

\section{Summary and conclusions }

We have shown that the anisotropic UKL Hamiltonian can be derived from the $S=1$ Kondo lattice model, if a local anisotropy term is included. A large variety of possible phase diagrams has been obtained, exhibiting interesting behaviors, as the possibility of abrupt transitions in the order parameters or re-entrant Kondo and FM phases at finite temperatures. The results have been shown for two different values of the parameter $P$, which measures the delocalization of the $5f$ electrons. The changes when the calculations are performed with other possible values of $P$ are only qualitative.

Finally, the results shown in this paper present some similarities with experimental results on two groups of compounds. For $P=0.20$, the variation of anisotropy can describe the variation of the Kondo and Curie temperatures as a function of pressure or magnetic field in UGe$_2$~\cite{Saxena2000,Hardy2009,Taufour2010,Troc2012}. In some regions of the phase diagram, a transition between two different ferromagnetic orders is observed, with a higher magnetic moment for $T\rightarrow0$. In our curves this can be observed in figure \ref{fig:pa_p020_p030}({\bf b}), where the transition occurs from FM to FM+Kondo phases. For $P=0.30$, the kink found in \ref{fig:pa_p020_p030}({\bf e}) is also observed in some neptunium compounds (NpNiSi$_2$, Np$_2$PdGa$_3$ and Np$_2$PtGa$_3$)~\cite{Colineau2008,Tran2010,Tran2014}. In these compounds, the magnetization at small applied magnetic field collapses at low temperatures, while the magnetization grows to a finite value and goes trough a second order transition for higher temperature. 
So, the UKL model, and the idea of a possible coexistence of Kondo effect and ferromagnetic order, first suggested by B. Coqblin~\cite{Lehur1997,Coqblin1999,NunezRegueiro2004}, is very promising for the description of actinide compounds that present Kondo effect and ferromagnetism.

\

\begin{flushleft}
{\bf Acknowledgments}
\end{flushleft}

CT acknowledge the financial support of Capes-Cofecub Ph 743-12. This research was also supported in part by the Brazilian Ministry of Science, Technology and Innovation (MCTI) and the Conselho Nacional de Desenvolvimento Cient\'ifico e Tecnol\'ogico (CNPq). Research carried out with the aid of the Computer System of High Performance of the International Institute of Physics-UFRN, Natal, Brazil.

\

\begin{flushleft}
{\Large \bf Appendix}
\end{flushleft}

\appendix
\section{Schrieffer-Wolff transformation}
\label{sec:ap_sc}

The Schrieffer-Wolff transformation~\cite{Schrieffer1966} relates the Anderson and Kondo Hamiltonians. In this work we use the same method described in Ref.~\cite{Schrieffer1967}, where the canonical transformation permit us to obtain from  the hybridization term $H_{hyb}$, a new term $\tilde{H}$, give by
{\small
\begin{align}\label{eqcano_a}
\ez{b}\tilde{H}\ze{a}=\frac{1}{2}\sum_{c}\ez{b}H_{hyb}\ze{c}\ez{c}H_{hyb}\ze{a}
\left(\frac{1}{E_a-E_c}+\frac{1}{E_b-E_c}\right)\,.
\end{align}}
This equation represents the scattering process from an initial state $\ze{a}$, through a intermediate state $\ze{c}$, to a final state $\ze{b}$ where $E_a$, $E_c$ and $E_b$ are the respective eigenenergies. The states $\ze{a}$, $\ze{b}$ and $\ze{c}$ should be eigenstates of $H_0$. We basically consider two scattering process: the first one is the process where the initial and final states have one conduction electron and two $f$-electrons; the second scattering process have three $f$-electrons in both initial and final states.

The eigenstates and eigenenergies for the first process are defined as
\begin{align}
&\ckm{k}{\sa}\fim{i}{1}{\sa}\fim{i}{2}{\sa}\zer;\quad E=U'-J+\epk{k}+2\epf-D\,,\\
&\ckm{k}{\sa}\frac{1}{\sqrt{2}}\mzer;\notag \\
&\qquad E=U'-J+\epk{k}+2\epf\,,\\
&\ckm{k}{\sa}\ckm{k'}{\sa'}\fim{i}{\al}{\sa''}\zer;\quad E=\epk{k}+\epk{k'}+\epf-D/4\,.
\end{align}
For the second process we have
\begin{align}
&\fim{j}{\be}{\sa'}\fim{i}{1}{\sa}\fim{i}{2}{\sa}\zer;\quad E=U'-J+3\epf-5D/4\,,\\
&\fim{j}{\be}{\sa}\frac{1}{\sqrt{2}}\mzer; \\
&\qquad E=U'-J+3\epf-D/4\,,\\
&\ckm{k}{\sa}\fim{j}{\be}{\sa'}\fim{i}{\al}{\sa''}\zer;\quad E=2\epf+\epk{k}-D/2\,.&
\end{align}

The first scattering process gives origin to the Kondo interaction. The inclusion of anisotropy in the model, equation \ref{eq:UAL}, modifies the diagonal and the non-diagonal part of the Kondo interaction as shown below
\begin{align}\label{eq:kondo_a}
H_K=\frac{1}{2}\sum_{i\vk\vk'}&
\left[{J_{\vk,\vk'}}_{\perp}(\ckm{k'}{\up}\ck{k}{\pu}S_i^{f-}+\ckm{k'}{\pu}\ck{k}{\up}S_i^{f+})\right. \notag \\
&\left.+{J_{\vk,\vk'}}_{\parallel}(\ckm{k'}{\up}\ck{k}{\up}-\ckm{k'}{\pu}\ck{k}{\pu})S_{zi}^{f}\right]\,,
\end{align}
where the perpendicular and parallel part of $J_{\vk,\vk'}$ are
\begin{align}
&{J_{\vk,\vk'}}_{\perp}=-\vko{k'}{\al}\vka{k}{\al}e^{i(\vk-\vk')\cdot\vri}\times \notag \\ &\Big(\frac{1}{U'-J+\epf-\epk{k'}-3D/4}
+\frac{1}{U'-J+\epf-\epk{k}+D/4}\Big)\,,
\end{align}
\begin{align}
&{J_{\vk,\vk'}}_{\parallel}=-\vko{k'}{\al}\vka{k}{\al}e^{i(\vk-\vk')\cdot\vri}\times \notag \\ &\Big(\frac{1}{U'-J+\epf-\epk{k'}-3D/4}
+\frac{1}{U'-J+\epf-\epk{k}-3D/4}\Big).
\end{align}

\section{Mean field approach}
\label{sec:meanfield}
Taking into account all terms from the Schrieffer-Wolff transformation and
the terms from $H_0$ and $H_H$, a new Hamiltonian can be written as
\begin{align}
H\equiv H_0+H_H+\tilde{H}\,.
\end{align}
After a mean-field approximation an effective Hamiltonian is obtained and it reads 
\begin{align}
H=&\sum_{i\al\sa} E^f_{\sa}n_{i\al\sa}^f+\sum_{\vk\sa}\epsilon_{\vk\sa}n_{\vk\sa}^c+\sum_{\vk\al\sa}\Lambda_{\sa}\left(\lam{\vk}{\al}{\sa}+\lan{\vk}{\al}{\sa}\right)\notag\\
&+\sum_{\vk\al\sa}A_{\sa}\epk{k}\fim{\vk}{\al}{\sa}\fis{\vk}{\al}{\sa}+\mathbb{C}\,,
\end{align}
where
\begin{align}\label{eq:param_ani_a}
E_{\sa}^f&=\epf+U'\med{n_{\so}^f}+(U'-J)\med{n_{\sa}^f}+J_{\parallel}\sa\med{m^c}
&\notag \\&-\frac{J_{\parallel}}{8}\big(\med{\lambda_{\up}}^2+\med{\lambda_{\pu}}^2\big)
-\frac{J_{\perp}}{4}\med{\lambda_{\up}}\med{\lambda_{\pu}}
\nonumber\\&\quad
+J_Hz\sa\med{M^f}+B_{\sa}\,,\\
B_{\sa}&=-P\med{n_{\sa}^f\epsilon}\left[\frac{J_{\parallel}}{2}(1+2\sa\med{M^f})+\frac{J_{\perp}}{8}(1-2\sa\med{M^f})\right]\nonumber\\&+P\med{n_{\so}^f\epsilon}\left[\frac{J_{\parallel}}{8}(1+2\sa\med{M^f})-\frac{J_{\perp}}{8}(3+2\sa\med{M^f})\right],\\
\epsilon_{\vk\sa}&=\epk{k}+J_{\parallel}\sa\med{M^f}\,,\quad \sa=\pm\frac{1}{2}\,,\\
\Lambda_{\sa}&=-\frac{1}{4}\big(J_{\parallel}\med{\lambda_{\sa}}+J_{\perp}\med{\lambda_{\so}}\big)\,,\\
A_{\sa}&=-\frac{P}{32}\left[3J_{\parallel}+4J_{\perp}+4\sa\med{M^f}\left(5J_{\parallel}-2J_{\perp}\right)\right.\notag\\&\left.+3J_{\parallel}\med{M^f}^2\right]\,,\label{eq:aks}\\
\mathbb{C}&=-2U'N \med{n_{\up}^f}\med{n_{\pu}^f}-(U'-J)N\left(\med{n_{\up}^f}^2+\med{n_{\pu}^f}^2\right)
\notag\\&
+\frac{J_{\parallel}}{2}N\left(\lws{\up}^2+\lws{\pu}^2\right)+J_{\perp}N\lws{\up}\lws{\pu}
\notag\\&
-\frac{J_H}{2}zN\med{M^f}^2-J_{\parallel}N\med{m^c}\med{M^f}-4A_{\sa}\med{n_{\sa}^f\epsilon}\,,
\end{align}
The value of $zJ_H$ is renormalized by $zJ_H\rightarrow zJ_H+D$ ($z$ is the number of first neighbor and is taking equal to 6 in a cubic lattice). 
We consider the mean field parameters in the uniform solution, and we assume that both orbitals remain equivalent. We also define the average occupation number of conduction electrons, $\med{n^c}$, where the magnetization is $\med{m^c}=\frac{1}{2}(\med{n_{\up}^c}-\med{n_{\pu}^c})$, and the average occupation number of $f$ electrons per orbital, $\med{n^f}$.

The mean field parameters are obtained by calculating the Green's functions. Considering that the conduction electrons have a constant density of state $\rho_0=1/2W_D$, the self-consistent equations are written as: 

\begin{align}
\med{n_{\sa}^c} &=\rho_0\int_{-W_D}^{W_D}d\epsilon\Big[F_{1\sa}(\epsilon)-(A_{\sa}\epsilon+E_{\sa}^f)F_{2\sa}(\epsilon)\Big]\,,\\
\med{n_{\sa}^f} &=\frac{\rho_0}{2}\int_{-W_D}^{W_D}d\epsilon\Big[f(E_{\sa}^f+A_{\sa}\epsilon)+F_{1\sa}(\epsilon)\notag\\
&\quad-\big(\epsilon-\mu+\sa J_{\parallel}\med{M^f}\big)F_{2\sa}\Big]\,, \\
\med{\lambda_{\sa}} &=2\rho_0\int_{-W_D}^{W_D}F_{2\sa}(\epsilon)\Lambda_{\so}\,,\\
\med{n_{\sa}^f\epsilon}&=\frac{\rho_0}{2}\int_{-W_D}^{W_D}\epsilon d\epsilon\Big[f(E_{\sa}^f+A_{\sa}\epsilon)+F_{1\sa}(\epsilon)\notag\\
&\quad-\big(\epsilon-\mu+\sa J_{\parallel}\med{M^f}\big)F_{2\sa}\Big]\,,
\end{align}
where $\mu$ is the chemical potential, $f(\omega)=(1+e^{\omega/T})^{-1}$ is the Fermi-Dirac distribution and
\begin{align}
F_{1\sa}(\epsilon)&=f[\Omega_{\sa}^+(\epsilon)]-f[\Omega_{\sa}^-(\epsilon)] \\
F_{2\sa}(\epsilon)&=\frac{f[\Omega_{\sa}^+(\epsilon)]-f[\Omega_{\sa}^-(\epsilon)]}{\Delta\Omega_{\sa}(\epsilon)}\,,
\end{align}
with
\begin{align}
\Omega_{\sa}^{\pm}(\epsilon)&=\frac{1}{2}\Big[\epsilon(1+A_{\sa})-\mu+E_{\sa}^f+\sa J_{\parallel}\med{M^f}\pm\Delta\Omega_{\sa}(\epsilon)\Big]\,,\\
\Delta\Omega_{\sa}(\epsilon)&=\sqrt{
\big[\epsilon(1-A_{\sa})-\mu-E_{\sa}^f+\sa J_{\parallel}\med{M^f}\big]^2+8(\Lambda_{\sa})^2
}\,.
\end{align}

\bibliographystyle{nature}
\bibliography{reference_140428}

\begin{thebibliography}{10}

\bibitem{Doniach1977}
Doniach, S.
\newblock {Phase diagram for the {K}ondo lattice}.
\newblock In Procceding of the International Conference on Valence
  Instabilities and Related Narrow-Band Phenomena, {R. D. Parks}, editor,  169
  (Plenum Press, New York, 1977).

\bibitem{Iglesias1997}
Iglesias, J.~R., Lacroix, C., and Coqblin, B.
\newblock {Revisited Doniach diagram: Influence of short-range
  antiferromagnetic correlations in the Kondo lattice}.
\newblock {\em Physical Review B}{ \bf 56}(18), 820--826 (1997).

\bibitem{Nozieres1980}
Nozi\`{e}res, P. and Blandin, A.
\newblock {Kondo effect in real metals}.
\newblock {\em Journal de Physique}{ \bf 41}(3), 193--211 (1980).

\bibitem{Schoenes1996}
Schoenes, J., Vogt, O., L\"{o}hle, J., Hulliger, F., and Mattenberger, K.
\newblock {Variation of f-electron localization in diluted US and UTe.}
\newblock {\em Physical Review B}{ \bf 53}(22), 14987--14995, June  (1996).

\bibitem{Bukowski2005}
Bukowski, Z., Tro\'{c}, R., Stepień-Damm, J., Sulkkowski, C., and Tran, V.~H.
\newblock {Single-crystalline study of the ferromagnetic Kondo}.
\newblock {\em Journal of Alloys and Compounds}{ \bf 403}, 65--70 (2005).

\bibitem{Colineau2008}
Colineau, E., Wastin, F., Sanchez, J.~P., and Rebizant, J.
\newblock {Magnetic properties of NpNiSi$_2$}.
\newblock {\em Journal of Physics: Condensed Matter}{ \bf 20}(7), 075207,
  February  (2008).

\bibitem{Tran2010}
Tran, V.~H., Griveau, J.-C., Eloirdi, R., Miiller, W., and Colineau, E.
\newblock {Magnetic and electronic properties of the ferromagnetic
  Kondo-lattice system Np$_2$PdGa$_3$}.
\newblock {\em Physical Review B}{ \bf 82}(9), 1--11, September  (2010).

\bibitem{Tran2014}
Tran, V.~H., Griveau, J.-C., Eloirdi, R., and Colineau, E.
\newblock {Ferromagnetic behavior of the Kondo lattice compound
  Np$_2$PtGa$_3$}.
\newblock {\em Physical Review B}{ \bf 89}, 054424, Feb  (2014).

\bibitem{Zwicknagl2002}
Zwicknagl, G., Yaresko, A., and Fulde, P.
\newblock {Microscopic description of origin of heavy quasiparticles in
  UPt$_3$}.
\newblock {\em Physical Review B}{ \bf 65}(8), 081103(R), February  (2002).

\bibitem{Zwicknagl2003}
Zwicknagl, G., Yaresko, A., and Fulde, P.
\newblock {Fermi surface and heavy masses for UPd$_2$Al$_3$}.
\newblock {\em Physical Review B}{ \bf 68}(5), 052508, August  (2003).

\bibitem{Hoshino2013}
Hoshino, S. and Kuramoto, Y.
\newblock {Itinerant Versus Localized Heavy-Electron Magnetism}.
\newblock {\em Physical Review Letters}{ \bf 111}(2), 026401, July  (2013).

\bibitem{Schoenes1984}
Schoenes, J., Frick, B., and Vogt, O.
\newblock {Transport properties of uranium monochalcogenide and monopnictide
  single crystals}.
\newblock {\em Physical Review B}{ \bf 30}(11), 6578 (1984).

\bibitem{Sheng1996}
Sheng, Q.~G. and Cooper, B.~R.
\newblock {Pressure-induced magnetic ordering effects in correlated-electron
  uranium monochalcogenides}.
\newblock {\em Journal of Magnetism and Magnetic Materials}{ \bf 164}(3),
  335--344, December  (1996).

\bibitem{Thomas2011}
Thomas, C., Sim\~{o}es, A. S.~R., Iglesias, J.~R., Lacroix, C., Perkins, N.,
  and Coqblin, B.
\newblock {Application of the S=1 underscreened Anderson lattice model to Kondo
  uranium and neptunium compounds}.
\newblock {\em Physical Review B}{ \bf 83}(1), 125102, January  (2011).

\bibitem{Perkins2007}
Perkins, N.~B., N\'{u}\~{n}ez Regueiro, M.~D., Coqblin, B., and Iglesias, J.~R.
\newblock {The underscreened Kondo lattice model applied to heavy fermion
  uranium compounds}.
\newblock {\em Physical Review B}{ \bf 76}, 125101 (2007).

\bibitem{Thomas2012}
Thomas, C., Sim\~{o}es, A. S.~R., Iglesias, J.~R., Lacroix, C., and Coqblin, B.
\newblock {Application of the underscreened Kondo lattice model to neptunium
  compounds}.
\newblock {\em Journal of Physics: Conference Series}{ \bf 391}, 012174,
  December  (2012).

\bibitem{Cornut1972}
Cornut, B. and Coqblin, B.
\newblock {Influence of Crystalline Field On Kondo Effect of Alloys and
  Compounds With Cerium Impurities}.
\newblock {\em Physical Review B}{ \bf 5}(11), 4541 (1972).

\bibitem{Coqblin2006}
Coqblin, B.
\newblock {Strongly Correlated Electron Behaviors and Heavy Fermions in
  Anomalous Rare-earth and actinide Systems}.
\newblock In Lectures on the Physics of Highly Correlated Electron Systems X,
  Avella, A. and Mancini, F., editors, volume 846, ~3. AIP Conference
  Proceedings,  (2006).
\newblock Tenth Training Course in the Physics of Correlated Electron Systems
  and High Tc Superconductors.

\bibitem{Schrieffer1966}
Schrieffer, J.~R. and Wolff, P.~A.
\newblock {Relation between Anderson and Kondo Hamiltonians}.
\newblock {\em Physical Review}{ \bf 149}(2), 491 (1966).

\bibitem{Zhang2000}
Zhang, G.-M. and Yu, L.
\newblock {Kondo singlet state coexisting with antiferromagnetic long-range
  order: A possible ground state for Kondo insulators}.
\newblock {\em Physical Review B}{ \bf 62}, 76--79, Jul  (2000).

\bibitem{Link1992}
Link, P., Benedict, U., Wittig, J., and W\"{u}hl, H.
\newblock {High-pressure resistance study of UTe}.
\newblock {\em Journal of Physics: Condensed Matter}{ \bf 4}(25), 5585--5589,
  June  (1992).

\bibitem{Cornelius1996}
Cornelius, A.~L., Schilling, J.~S., Vogt, O., Mattenberger, K., and Benedict,
  U.
\newblock {High-pressure susceptibility studies on the ferromagnetic uranium
  monochalcogenides US, USe and UTe}.
\newblock {\em Journal of Magnetism and Magnetic Materials}{ \bf 161},
  169--176, August  (1996).

\bibitem{Cooper1998}
Cooper, B.~R. and Lin, Y.-L.
\newblock {$f$-Electron delocalization/localization and the abrupt
  disappearance of uranium magnetic ordering with dilution alloying}.
\newblock {\em Journal of Applied Physics}{ \bf 83}(11), 6432 (1998).

\bibitem{Saxena2000}
Saxena, S.~S., Agarwal, P., Ahilan, K., Grosche, F.~M., Haselwimmer, R.~K.,
  Steiner, M.~J., Pugh, E., Walker, I.~R., Julian, S.~R., Monthoux, P.,
  Lonzarich, G.~G., Huxley, A., Sheikin, I., Braithwaite, D., and Flouquet, J.
\newblock {Superconductivity on the border of itinerant-electron ferromagnetism
  in UGe$_2$}.
\newblock {\em Nature}{ \bf 406}(6796), 587--92, August  (2000).

\bibitem{Hardy2009}
Hardy, F., Meingast, C., Taufour, V., Flouquet, J., v.~L\"{o}hneysen, H.,
  Fisher, R.~A., Phillips, N.~E., Huxley, A., and Lashley, J.~C.
\newblock {Two magnetic Gr\"{u}neisen parameters in the ferromagnetic
  superconductor UGe$_2$}.
\newblock {\em Physical Review B}{ \bf 80}(17), 174521, November  (2009).

\bibitem{Taufour2010}
Taufour, V., Aoki, D., Knebel, G., and Flouquet, J.
\newblock {Tricritical Point and Wing Structure in the Itinerant Ferromagnet
  UGe$_2$}.
\newblock {\em Physical Review Letters}{ \bf 105}(21), 217201, November
  (2010).

\bibitem{Troc2012}
Tro\'{c}, R., Gajek, Z., and Pikul, A.
\newblock {Dualism of the $5f$ electrons of the ferromagnetic superconductor
  UGe$_2$ as seen in magnetic, transport, and specific-heat data}.
\newblock {\em Physical Review B}{ \bf 86}(22), 224403, December  (2012).

\bibitem{Lehur1997}
{Le Hur}, K. and Coqblin, B.
\newblock {Underscreened Kondo effect: A two $S=1$ impurity model}.
\newblock {\em Physical Review B}{ \bf 56}(2), 668--677, July  (1997).

\bibitem{Coqblin1999}
Coqblin, B., Bernhard, B.-H., Iglesias, J.~R., Lacroix, C., and Le~Hur, K.
\newblock {Narrow-band effects in rare-earths and actinides: Interaction
  between the Kondo effect and magnetism}.
\newblock {\em Electron Correlations and Materials Properties}{ \bf 1}, 225
  (1999).

\bibitem{NunezRegueiro2004}
N\'u\~nez Regueiro, M.~D., Coqblin, B., and Iglesias, J.~R.
\newblock {The underscreened Kondo lattice model: application to uranium
  compounds}.
\newblock {\em Journal of Magnetism and Magnetic Materials}{ \bf 272},
  E95--E96, May  (2004).

\bibitem{Schrieffer1967}
Schrieffer, J.~R.
\newblock {Kondo effect - link between magnetic and nonmagnetic impurities in
  metals?}
\newblock {\em Journal of Applied Physics}{ \bf 38}(3), 1143 (1967).

\end{thebibliography}

\end{multicols}
 
\end{document}